\documentclass[conference]{IEEEtran}
\IEEEoverridecommandlockouts
\usepackage[letterpaper, left=0.62in, right=0.62in, bottom=1in, top=0.75in]{geometry}
\usepackage{amssymb,amsmath}
\usepackage{graphicx}
\usepackage{cite}
\usepackage{txfonts}
\usepackage{mathrsfs}
\usepackage{fontenc}
\usepackage[binary-units]{siunitx}
\DeclareSIUnit{\bps}{bps}

\usepackage[caption=false,font=footnotesize]{subfig}
\usepackage{tabularx}
\usepackage{multirow}
\usepackage{diagbox}
\usepackage{algorithm2e}
\usepackage{bbm}
\usepackage{xcolor}

\begin{document}
\title{Capacity Analysis and Rate Maximization Design in RIS-Aided Uplink Multi-User MIMO}

\author{\IEEEauthorblockN{Wei Jiang\IEEEauthorrefmark{1}\IEEEauthorrefmark{2} and Hans D. Schotten\IEEEauthorrefmark{2}\IEEEauthorrefmark{1}}
\IEEEauthorblockA{\IEEEauthorrefmark{1}Intelligent Networking Research Group, German Research Center for Artificial Intelligence (DFKI),  Germany\\
  }
\IEEEauthorblockA{\IEEEauthorrefmark{2}Department of Electrical and Computer Engineering, Technische Universit\"at (TU) Kaiserslautern, Germany\\
 }

}

\maketitle

\begin{abstract}
Reconfigurable intelligent surface (RIS) has recently drawn intensive attention due to its potential of simultaneously realizing high spectral and energy efficiency in a sustainable way. This paper focuses on the design of efficient transmission methods to maximize the uplink sum throughput in a RIS-aided multi-user multi-input multi-output (MU-MIMO) system. To provide an insightful basis, the channel capacity of RIS-aided MU-MIMO is theoretically analyzed. Then, the conventional transmission schemes based on orthogonal multiple access are presented as the baseline. From the information-theoretic perspective, we propose two novel schemes, i.e., \textit{joint transmission} based on the semidefinite relaxation of quadratic optimization problems and \textit{opportunistic transmission} relying on the best user selection. The superiority of the proposed schemes over the conventional ones in terms of achievable rates is justified through simulation results.
\end{abstract}

\IEEEpeerreviewmaketitle

\section{Introduction}

Reconfigurable intelligent surface (RIS) has recently attracted intensive attention from academia and industry \cite{Ref_renzo2020smart}. Through smartly adjusting the reflection coefficients of a large number of reconfigurable elements over a planar meta-surface \cite{Ref_wu2020towards}, an on-demand propagation environment is achieved for signal amplification or interference suppression, so as to improve the performance of wireless communications. Since the reflecting elements are nearly passive, low-cost, and lightweight, the RIS is a green and cost-efficient technology. It enables sustainable capacity and performance growth for  legacy 5G networks and the forthcoming 6G system \cite{Ref_jiang2021road, Ref_jiang20226G}.

Prior works on RIS-aided communications mostly focus on point-to-point communications that consider a base station (BS), a surface, and a single user. Depending on the number of antennas, the research works span from single-input single-output (SISO) to multi-input multi-output (MIMO). Nevertheless, a practical wireless system needs to accommodate many users simultaneously, imposing the necessity of studying multi-user MIMO (MU-MIMO). There has been a few recent works on this topic.  The authors of \cite{Ref_yan2020passive} developed a novel technique for passive beamforming and information transfer in RIS-aided MU-MIMO systems. In \cite{Ref_you2020energy}, a trade-off between energy and spectral efficiency in MU-MIMO uplink communications aided by a discrete-phase-shift RIS is discussed. The design of linear or nonlinear receivers for MU-MIMO systems aided by multiple RISs is studied in \cite{Ref_lv2021multiuser}. Zheng \textit{et al.} aimed to unveil the full potential of multi-RIS assisted wireless networks by studying a double-RIS multi-user communication system with cooperative passive beamforming in \cite{Ref_zheng2021doubleIRS}.  The work \cite{Ref_you2021reconfigurable} jointly optimizes the uplink transmit beamforming and the phase-shift matrix to maximize the system energy efficiency under partial channel state information (CSI). In \cite{Ref_niu2022double}, the effect of double RISs in improving the spectral efficiency of an MU-MIMO network operating in millimeter wave  is investigated. Joint beamforming and modulation design for embedding extra data into carrier signals from the BS to the RIS in a downlink MU-MIMO network is proposed in \cite{Ref_rehman2022modulating}. The work \cite{Ref_hu2022reconfigurable} presents a novel symbiotic radio system on the basis of RIS-aided MU-MIMO to enhance the primary transmission and simultaneously transmit its own information by back-scattering modulation. In addition, some other works such as \cite{Ref_liu2020matrix, Ref_zheng2021efficient} focus on one of the fundamental challenges, namely the acquisition of cascaded channel information, in RIS-assisted MU-MIMO systems.

This paper focuses on designing efficient transmission for a RIS-aided MU-MIMO system with the aim of maximizing its uplink sum throughput. To provide an insightful basis, an information-theoretic analysis in terms of the sum capacity is theoretically conducted. The conventional orthogonal multiple access (OMA) schemes, including time-division multiple access (TDMA) and frequency-division multiple access (FDMA), are presented as the baseline. Then, we propose two novel schemes, i.e., \textit{joint transmission (JT)} based on the semidefinite relaxation of quadratic optimization problems and \textit{opportunistic transmission (OT)} relying on the best user selection. The superiority of the proposed schemes over OMA in terms of achievable sum rate is justified through Monte-Carlo simulation.

The rest of the paper is organized as follows: Section II introduces the system model. Section III analyzes the channel capacity. In Section IV, the proposed JT and OT schemes are elaborated in comparison with the OMA schemes. Simulation setup and numerical results are demonstrated in Section V. Finally, Section VI concludes this paper.

\section{System Model}
Consider a RIS-aided multi-user MIMO communications system, which comprises an $N_b$-antenna BS, $K$ single-antenna user equipment (UE), and a surface with $N_s$ reconfigurable elements \cite{Ref_wu2019intelligent}. MU-MIMO is an \textit{asymmetric} system, where the downlink from a BS to several UEs is referred to as \textit{Gaussian MIMO broadcast channel}, while the uplink from multiple UEs to the BS is called \textit{Gaussian MIMO multiple access channel}.
This paper merely focuses on the uplink transmission while its analysis and development also provide some meaningful insights on the downlink transmission.

\begin{figure}[!t]
    \centering
    \includegraphics[width=0.38\textwidth]{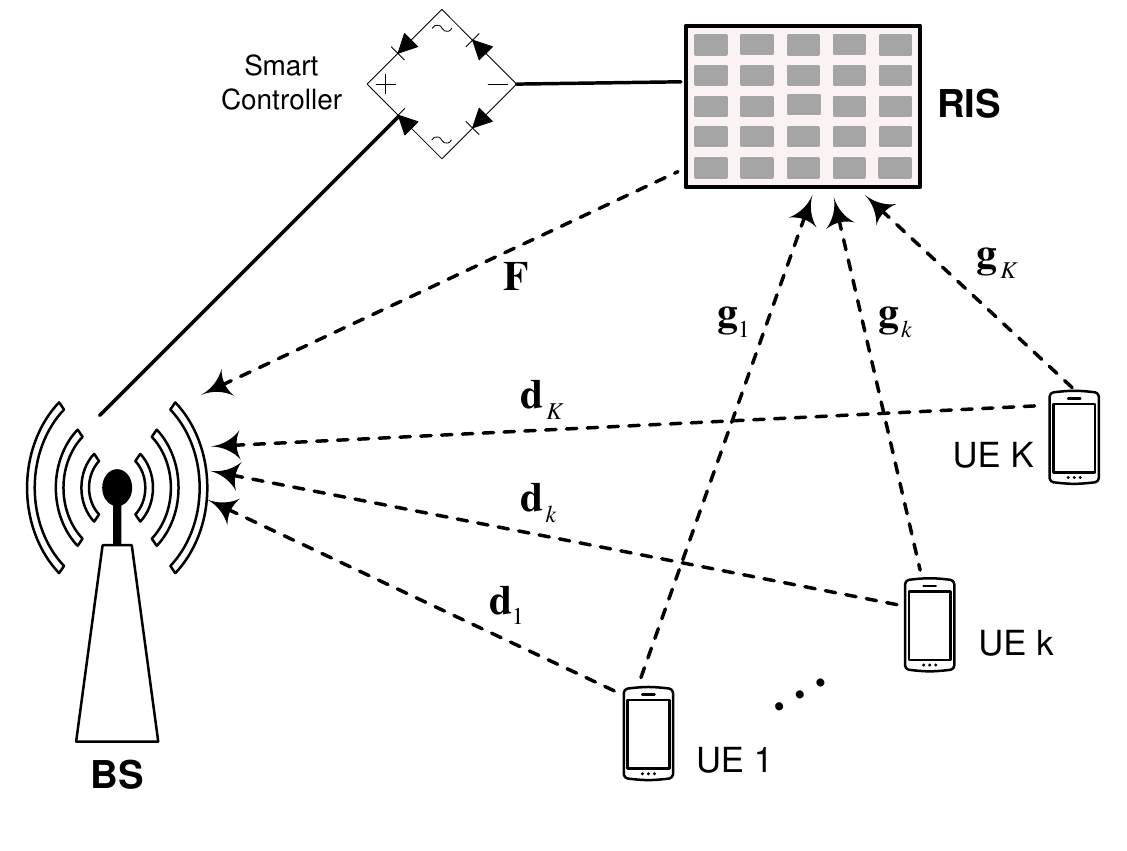}
    \caption{Schematic diagram of the uplink transmission of a RIS-aided MU-MIMO system consisting of a BS, a RIS, and $K$ users.  }
    \label{fig:SystemModel}
\end{figure}

As demonstrated in \figurename \ref{fig:SystemModel}, multiple UEs simultaneously send its respective signal towards the BS over the same time-frequency resource. The BS acquires  the uplink instantaneous CSI through estimating the pilot signals during the uplink training period. To facilitate the theoretical analysis, the BS is assumed to perfectly know the  CSI of all involved channels, as  prior works \cite{Ref_yan2020passive}-\cite{Ref_hu2022reconfigurable}.
The RIS is equipped with a smart controller that adaptively adjusts the phase shift of each reflecting element according to the knowledge of CSI.  Mathematically, a typical element $n\in \{1,2,\ldots,N_s\}$ is modeled by a reflection coefficient $\epsilon_{n}=a_{n} e^{j\theta_{n}}$, where $\theta_{n}\in [0,2\pi)$ denotes an induced phase shift, and $a_{n}\in [0,1]$ stands for amplitude attenuation. Although the practical RIS implementation supports a finite number of discrete phase shifts,  only a few  phase-control bits (e.g., $2$ bits as illustrated in \cite{Ref_jiang2022multiuser}) are sufficient for achieving near-optimal performance as continuous phase shifts. Without loss of generality, we use continuous phase shifts hereinafter for simplicity. As mentioned by \cite{Ref_wu2019intelligent}, $a_{n}=1$, $\forall n$ is the optimal setting that maximizes the signal strength and simplifies the implementation. Therefore, the RIS optimization focuses on a diagonal phase-shift matrix defined as $\boldsymbol{\Phi}=\mathrm{diag}\{e^{j\theta_{1}},\ldots,e^{j\theta_{N_s}}\}$.

 We use $s_k\in \mathbb{C}$ to denote the information symbol from user $k$, satisfying $\mathbb{E}[|s_k|^2]\leqslant P_k$, where $P_k$ denotes the power constraint of user $k$. All information symbols form a transmitted vector $\textbf{s}\in \mathbb{C}^{ K\times 1}=[s_1,s_2,\ldots,s_K]^T$.
Let $\textbf{D} \in \mathbb{C}^{N_b\times K}$ denote the channel matrix from $K$ users to $N_b$ receive antennae at the BS, and $\textbf{d}_k \in \mathbb{C}^{N_b\times 1}$ denotes the spatial signature of user $k$ impinged on the BS antenna array, we have $\mathbf{D}=\left[\mathbf{d}_{1},\mathbf{d}_{2}, \ldots, \mathbf{d}_{K}\right]$. Let $\textbf{G} \in \mathbb{C}^{N_s\times K}$ denote the channel matrix from $K$ users to $N_s$ reflecting elements, we have $\mathbf{G}=\left[\mathbf{g}_{1},\mathbf{g}_{2}, \ldots, \mathbf{g}_{K}\right]$, where $\textbf{g}_k \in \mathbb{C}^{N_s\times 1}$ represents the spatial signature of user $k$ impinged over the RIS. Similarly, we write $\textbf{F} \in \mathbb{C}^{N_b\times N_s}$ to denote the channel matrix from the RIS to the BS. Without losing generality, any entry within these channel vectors or matrices is modeled as
a circularly symmetric complex Gaussian random variable denoted by $X\sim \mathcal{CN}(\mu,\sigma_c^2)$, where $\mu$ denotes the mean, and $\sigma_c^2$ is the average channel (power) gain.

The overall system can be modelled as
\begin{equation} \label{eqn:MUMIMO:BSchannelmodel}
    \mathbf{y} = \biggl(\mathbf{F} \boldsymbol{\Phi} \mathbf{G} +\mathbf{D}\biggr) \textbf{s}+\textbf{n},
\end{equation}
where $\textbf{y}=[y_1,y_2,\ldots,y_{N_b}]^T$ stands for the received vector, and $\textbf{n}\in \mathcal{CN}(\mathbf{0},\sigma_n^2\mathbf{I}_{N_b})$ is independent and identically distributed (\textit{i.i.d.}) additive white Gaussian noise (AWGN) with zero mean and variance $\sigma_n^2$.
Decomposing (\ref{eqn:MUMIMO:BSchannelmodel}), the signal model can be rewritten as an alternative form
\begin{equation}  \label{eqn_IRS_signalModel}
     \mathbf{y} = \sum_{k=1}^K \biggl( \mathbf{F} \boldsymbol{\Phi} \mathbf{g}_k +\mathbf{d}_{k}\biggr)s_k + \textbf{n}.
\end{equation}

\section{Capacity Analysis}

In a point-to-point system, the channel capacity provides a measure of the performance limit: reliable communications with an arbitrarily small error probability can be achieved at any rate $R<C$, whereas reliable communications are impossible when $R>C$. For a multi-user system consisting of a BS and $K$ UEs, the concept is extended to a similar performance metric called \textit{a capacity region} \cite{Ref_tse2005fundamentals}. It is characterized by a $K$-dimensional space $\mathfrak{C} \in \mathbb{R}_+^K$, where $\mathbb{R}_+$ denotes the set of non-negative real-valued numbers, and $\mathfrak{C}$ is the set of all K-tuples $(R_1,R_2,\ldots,R_{K})$ such that a generic user $k$ can reliably communicate at rate $R_k$ simultaneously with others. Due to the shared transmission resource, there is a trade-off: if one desires a higher rate, some of other users have to lower their rates. From this capacity region, a performance metric can be derived, i.e., the sum capacity
\begin{equation}
    C_{sum}= \max_{(R_1,R_2,\ldots,R_{K})\in \mathfrak{C}} \left( \sum_{k=1}^K R_k   \right),
\end{equation}
indicating the maximum total throughput that can be achieved.

The achievable rate of a typical user is limited by the \textit{single-user bound}, which is the capacity of the point-to-point link with the other users absent from the system. From \eqref{eqn_IRS_signalModel}, we have
\begin{equation}
  R_k < \log  \left[  1+\frac{\left\|\mathbf{F} \boldsymbol{\Phi} \mathbf{g}_k +\mathbf{d}_{k}\right\|^2P_k}{\sigma_n^2}  \right],\: \forall\: k.
\end{equation}
In addition, any combination of user rates are constrained by
\begin{align} \nonumber
  \sum_{k\in\mathcal{S}} R_k &< \log \det \left[  \mathbf{I}_{N_b}+\frac{ \sum_{k\in\mathcal{S}} \left\|\mathbf{F} \boldsymbol{\Phi} \mathbf{g}_k +\mathbf{d}_{k}\right\|^2P_k}{\sigma_n^2}  \right],\\
  &\forall \mathcal{S}\in \Bigl\{1,2,\ldots,K \Bigr\}.
\end{align}

For the ease of notation, we write $\mathbf{h}_k=\biggl( \mathbf{F}\boldsymbol{\Phi}\mathbf{g}_k  +\mathbf{d}_{k}\biggr)$ to denote the effective channel from user $k$ to the BS with the aid of the RIS.
The capacity region is now a $K$-dimensional polyhedron, which can be mathematically described by
\begin{equation}
  \mathfrak{C} = \left \{ (R_1,\ldots,R_K)\in \mathbb{R}_+^K  \left | \\ \begin{aligned}
    \sum_{k\in\mathcal{S}} R_k &{<} \log \det \left[  \mathbf{I}_{N_b}{+}\frac{ \sum_{k\in\mathcal{S}} \left\|\mathbf{h}_{k}\right\|^2P_k}{\sigma_n^2}  \right]\\
  &\forall \mathcal{S}\in \Bigl\{1,2,\ldots,K \Bigr\}\\
\end{aligned} \right. \right\}.
\end{equation}
The \textit{sum capacity} of a RIS-aided MU-MIMO system can be given by
\begin{equation} \label{EQN_SumRate}
    C_{sum}=   \log \det \left[  \mathbf{I}_{N_b}+\frac{ \sum_{k=1}^K \left\|\mathbf{F} \boldsymbol{\Phi} \mathbf{g}_k +\mathbf{d}_{k}\right\|^2P_k}{\sigma_n^2}  \right].
\end{equation}

It is not difficult to derive that the instantaneous channel gain of multiple users is equivalent to an overall channel gain, namely
\begin{equation}
    \sum_{k=1}^K \left\|\mathbf{F} \boldsymbol{\Phi} \mathbf{g}_k +\mathbf{d}_{k}\right\|^2=\left\|\mathbf{F} \boldsymbol{\Phi} \mathbf{G} +\mathbf{D}\right\|^2.
\end{equation}
If all users have the same power constraint, i.e., $P_k=P_u$, $\forall k$, the sum capacity in \eqref{EQN_SumRate} is rewritten as
\begin{equation} \label{eqn_IRS_sumRate}
    C_{sum}=   \log \det \left[  \mathbf{I}_{N_b}+\frac{  \left\|\mathbf{F} \boldsymbol{\Phi} \mathbf{G} +\mathbf{D} \right\|^2P_u}{\sigma_n^2}  \right].
\end{equation}

\section{Sum-Rate Maximization Design}
The aim of this paper is to design efficient transmission for the maximization of the sum rate in the uplink RIS-aided MU-MIMO systems.
From \eqref{eqn_IRS_sumRate}, the sum capacity is a function of $\boldsymbol{\Phi}$, resulting in the following optimization formula
\begin{equation}
\begin{aligned} \label{EQN:sumRate_Optimization}
\max_{\boldsymbol{\Phi}} \quad & \left\|\mathbf{F} \boldsymbol{\Phi} \mathbf{G} +\mathbf{D}\right\|^2\\
\textrm{s.t.} \quad & \theta_{n}\in [0,2\pi), \: \forall n=1,\ldots,N_s,
\end{aligned}
\end{equation}
which is mathematically intractable.

Fortunately, it is observed that the BS-RIS link generally has a strong line-of-sight (LOS) path since both nodes are stationary, and their deployment locations are deliberately selected without any blockage in-between. In contrast to randomly distributed, moving UEs, the BS-RIS channel exhibits high correlation and sparsity. Furthermore, if the BS applies a correlated antenna array, e.g., with a small inter-element spacing of half wavelength, the channel can be modelled as the product of the channel vector of the reference antenna and the steering vector of the array \cite{Ref_yang2013random}. Therefore, the BS-RIS link can be represented by the channel vector $\textbf{f} \in \mathbb{C}^{1\times N_s}$ between the reference antenna and the RIS. Accordingly, the direct channel is degraded from $\mathbf{D}$ to $\textbf{d} \in \mathbb{C}^{1\times K}$.
As a result, \eqref{EQN:sumRate_Optimization} is simplified to
\begin{equation}
\begin{aligned} \label{EQN_MU_IRS_optimizationformular}
\max_{\boldsymbol{\Phi}} \quad & \left\|\mathbf{f} \boldsymbol{\Phi} \mathbf{G} +\mathbf{d}\right\|^2\\
\textrm{s.t.} \quad & \theta_{n}\in [0,2\pi), \: \forall n=1,\ldots,N_s.
\end{aligned}
\end{equation}

The objective function in \eqref{EQN_MU_IRS_optimizationformular} becomes solvable since it is quadratically constrained quadratic program (QCQP) optimization \cite{Ref_sidiropoulos2006transmit}, based on which \textit{joint transmission} is proposed. In addition, \textit{opportunistic transmission} relying on the best user selection is also provided. For comparison, this section first presents the behaviours of the conventional OMA schemes including TDMA and FDMA in RIS-aided MU-MIMO systems.

\subsection{Orthogonal Multiple Access}
\subsubsection{TDMA-RIS}  It is a simple scheme by dividing the signaling dimension along the time axis into $K$ orthogonal slots.     Using the round-robin scheduling, each user cyclically accesses to its assigned slot. A general user $k$ transmits $s_k$ at the $k^{th}$ slot while other users keep silent. According to \cite{Ref_zhang2018space},  a RIS element made by positive-intrinsic-negative (PIN) diodes has a maximal switching frequency of \SI{5}{\mega\hertz}, much faster than the shifting of time slots typically on the order of millisecond (\si{\milli\second}). It implies that the phase-shift matrix can be adjusted per slot, denoted by $\boldsymbol{\Phi}_k=\mathrm{diag}\{e^{j\theta_{1}[k]},\ldots,e^{j\theta_{N_s}[k]}\}$, $k=1,\ldots,K$ with the time-selective phase shift $\theta_{n}[k]$. The received signal vector for user $k$ is given by \begin{equation}
    \mathbf{y}_k = \biggl(\mathbf{f} \boldsymbol{\Phi}_k \mathbf{g}_k +d_k\biggr) s_k+\textbf{n},
\end{equation} where $d_k$ is the $k^{th}$ element of $\mathbf{d}$. The phase of each reflected signal should be tuned to align with the phase of the LOS signal for coherent combining at the receiver. Thus, it is not hard to derive that the optimal phase-shift matrix equals
\begin{equation} \label{eqn_IRS_phaseshifts}
  \boldsymbol{\Phi}_k^\star =\mathrm{diag}\left \{e^{j\left(\arg\left(d_k\right)-\arg\left( \mathrm{diag}(\mathbf{f})\mathbf{g}_k \right)\right)} \right\},
\end{equation}
where $\arg(\cdot)$ stands for the phase of a complex scalar or vector. It results in a per-user rate of
\begin{equation}
    R_k=\frac{1}{K}\log \left(1+\left|\sum_{n=1}^{N_s}|f_n||g_{n,k}|+|d_k| \right|^2 \frac{P_u  }{\sigma_n^2}\right),
\end{equation} where $f_n$ denotes the channel coefficient between the BS and reflecting element $n$, and $g_{n,k}$ denotes the channel coefficient between reflecting element $n$ and user $k$.
Thereby, the sum rate of the TDMA-RIS system can be computed by
\begin{equation} \label{IRS_EQN_TDMA_SE}
    C_{tdma}=\sum_{k=1}^K\frac{1}{K}\log\left(1+\frac{P_u \Bigl|\sum_{n=1}^{N_s}|f_n||g_{n,k}|+|d_k| \Bigr|^2 }{\sigma_n^2} \right),
\end{equation}
where the factor $1/K$ is due to the orthogonal partitioning of the time resource.

\subsubsection{FDMA-RIS} The system bandwidth is split into $K$ orthogonal subchannels, and each user occupies a subchannel over the entire time. Unlike the \textit{time-selective} phase shifting in TDMA, the RIS is not \textit{frequency-selective} due to the hardware limitation. It implies that the surface can be optimized at most for a particular user, whereas other users suffer from phase-unaligned reflection. If the RIS aids the signal transmission of a dedicated user $\hat{k}$, the optimal phase-shift matrix $\boldsymbol{\Phi}_{\hat{k}}^\star$ can be obtained from \eqref{eqn_IRS_phaseshifts}. Its achievable sum rate is calculated by
 \begin{align} \nonumber \label{IRS_EQN_FDMA_SE}
     C_{fdma}&=\sum_{k=1}^K\frac{1}{K}\log\left(1+\frac{P_u \Bigl|\mathbf{f} \boldsymbol{\Phi}_{\hat{k}}^\star \mathbf{g}_k +d_k \Bigr|^2 }{\sigma_n^2} \right)\\
     &= \frac{1}{K}\log\left(1+\frac{P_u \Bigl|\sum_{n=1}^{N_s} |f_n||g_{n,{\hat{k}}}|+|d_{\hat{k}}| \Bigr|^2 }{\sigma_n^2} \right)  \\ \nonumber
     &+\sum_{k\neq {\hat{k}}} \frac{1}{K}\log\left(1+\frac{P_u \Bigl|\mathbf{f}\boldsymbol{\Phi}_{\hat{k}}^\star \mathbf{g}_k +d_{k}\Bigr|^2 }{\sigma_n^2} \right).
  \end{align}
Tuning the RIS to optimize different users yields different performance. The best user that maximizes the sum rate can be determined by exhaustively selecting each user as the target: \begin{equation}
        \hat{k}=\arg\max_{k\in \{1,2,\ldots,K\}} C_{fdma}.
    \end{equation}
In addition to the exhaustive search, the simplest way is to randomly select a user.

\subsection{Joint Transmission}
From the information-theoretic perspective, OMA is inefficient because each user utilizes only a fraction of the available time-frequency resource. With this regard, we propose a joint-transmission scheme for a RIS-aided MU-MIMO system, where all users transmit their signals simultaneously over the same time-frequency resource.

Unlike OMA, where the RIS is tuned for a particular user, the phase-shift matrix in JT needs to be optimized based on the CSI of all users.
Define $\mathbf{q}=\left[q_{1},q_{2},\ldots,q_{N_s}\right]^H$ with $q_{n}=e^{j\theta_{n}}$, $n=1,\ldots, N_s$ and  $\boldsymbol{\chi}=\mathrm{diag}(\mathbf{f})\mathbf{G}\in \mathbb{C}^{N_s\times K}$, we have $\mathbf{f} \boldsymbol{\Phi} \mathbf{G}=\mathbf{q}^H\boldsymbol{\chi}\in \mathbb{C}^{1\times K} $.
Thus, the objective function in \eqref{EQN_MU_IRS_optimizationformular} is transferred to $\left\|\mathbf{f} \boldsymbol{\Phi} \mathbf{G} +\mathbf{d}\right\|^2=\left\|\mathbf{q}^H\boldsymbol{\chi} +\mathbf{d}\right\|^2$, resulting in
\begin{equation}
\begin{aligned} \label{EQN_IRS_QCQPoptimization}
\max_{\mathbf{q}} \quad & \mathbf{q}^H\boldsymbol{\chi}\boldsymbol{\chi}^H\mathbf{q}+\mathbf{q}^H\boldsymbol{\chi}\mathbf{d}^H+\mathbf{d}\boldsymbol{\chi}^H\mathbf{q}+\|\mathbf{d}\|^2\\
\textrm{s.t.} \quad & |q_n|^2=1, \:  \forall n=1,\ldots,N_s,
\end{aligned}
\end{equation}
which is a non-convex QCQP problem \cite{Ref_sidiropoulos2006transmit}. Introducing an auxiliary variable $t$, \eqref{EQN_IRS_QCQPoptimization} can be homogenized as
\begin{equation} \begin{aligned} \label{eqn_IRS_relaxedOptimization}
    \max_{\mathbf{q}}  \quad & \left\|t\mathbf{q}^H\boldsymbol{\chi} +\mathbf{d}\right\|^2\\
     =\max_{\mathbf{q}} \quad & t^2\mathbf{q}^H\boldsymbol{\chi}\boldsymbol{\chi}^H\mathbf{q}+t\mathbf{q}^H\boldsymbol{\chi}\mathbf{d}^H+t\mathbf{d}\boldsymbol{\chi}^H\mathbf{q}+\|\mathbf{d}\|^2,
     \end{aligned}
\end{equation}

Defining
\begin{equation}
    \mathbf{C}=\begin{bmatrix}\boldsymbol{\chi}\boldsymbol{\chi}^H&\boldsymbol{\chi}\mathbf{d}^H\\ \mathbf{d}\boldsymbol{\chi}^H& \|\mathbf{d}\|^2\end{bmatrix},\:\:\mathbf{v}= \begin{bmatrix}\mathbf{q}\\ t\end{bmatrix},
\end{equation}
\eqref{eqn_IRS_relaxedOptimization} equals to
\begin{equation}
\begin{aligned} \label{EQN:Optimization}
\max_{\mathbf{v}} \quad & \mathbf{v}^H\mathbf{C}\mathbf{v}\\
\textrm{s.t.} \quad & |q_n|^2=1, \: \forall n=1,\ldots,N_s\\
\quad & |t|^2=1.
\end{aligned}
\end{equation}
Let $\mathbf{V}=\mathbf{v}\mathbf{v}^H$, we have $\mathbf{v}^H\mathbf{C}\mathbf{v}=\mathrm{Tr}(\mathbf{C}\mathbf{V})$, where $\mathrm{Tr}(\cdot)$ denotes the trace of a matrix.
As a result, \eqref{EQN:Optimization} is reformulated as
\begin{equation}  \label{RIS_EQN_optimizationTrace}
\begin{aligned} \max_{\mathbf{V}}\quad &  \mathrm{Tr} \left(  \mathbf{C}\mathbf{V} \right)\\
\textrm{s.t.}  \quad & \mathbf{V}_{n,n}=1, \: \forall n=1,\ldots,N_s \\
  \quad & \mathbf{V}\succ 1
\end{aligned},
\end{equation}
where $\mathbf{V}_{n,n}$ means the $n^{th}$ diagonal element of $\mathbf{V}$, and $\succ$ stands for a positive semi-definite matrix.
The optimization formula is transformed to a semi-definite program, whose globally optimal solution $\mathbf{V}^\star$ can be efficiently solved by available numerical algorithms such as CVX in MATLAB \cite{cvx}.

Conduct the eigenvalue decomposition $\mathbf{V}^\star=\mathbf{U}\boldsymbol{\Sigma}\mathbf{U}^H$, where $\mathbf{U}$ is a unitary matrix and $\boldsymbol{\Sigma}$ is a diagonal matrix, both with the size $(N_s+1)\times (N_s+1)$. A sub-optimal solution for the optimization problem is given by
\begin{equation}
    \bar{\mathbf{v}}=\mathbf{U}\boldsymbol{\Sigma}^{1/2}\mathbf{r},
\end{equation}
where $\mathbf{r}$ is a Gaussian random vector generated according to $\mathbf{r}\in \mathcal{CN}(\mathbf{0},\mathbf{I}_{N_s+1})$. Finally, the solution to the optimization problem  can be determined as
\begin{equation}
    \boldsymbol{\Phi}_{JT}^\star =\mathrm{diag}\left \{e^{j \arg\left( \left[\frac{\bar{\mathbf{v}}}{\bar{v}_{_{N_s+1}}} \right]_{1:N_s}\right)}\right\},
\end{equation}
where $[\cdot]_{1:N_s}$ denotes a sub-vector extracting the first $N_s$ elements, and $\bar{v}_{_{N_s+1}}$ is the last element of $\bar{\mathbf{v}}$.

Thus, the sum capacity of JT is computed as
\begin{equation}
    C_{JT}=\log\left(1+\frac{ \Bigl\| \mathbf{f} \boldsymbol{\Phi}_{JT}^\star \mathbf{G} +\mathbf{d} \Bigr\|^2 P_u}{\sigma_n^2} \right),
\end{equation}
where the factor $1/K$ in \eqref{IRS_EQN_TDMA_SE} and \eqref{IRS_EQN_FDMA_SE} is avoided due to the full exploitation of the time-frequency resource in JT. Ideally, the applied phase-shift matrix can optimally optimize the reflection for all users simultaneously, providing an upper performance bound of JT as
\begin{equation} \label{IRS_EQN_upper}
    C_{Upper}=\log\left(1+\frac{\sum_{k=1}^K \Bigl|\sum_{n=1}^{N_s}|f_n||g_{n,k}|+|d_k| \Bigr|^2 P_u}{\sigma_n^2} \right).
\end{equation}

The joint transmission for RIS-aided MU-MIMO systems is depicted also in Algorithm 1.
\SetKwComment{Comment}{/* }{ */}
\RestyleAlgo{ruled}
\begin{algorithm}
\caption{Joint RIS Transmission} \label{alg:IRS001}
\SetKwInput{kwInit}{Initialization}
\kwInit {$\mathbf{q}\gets \left[q_{1},q_{2},\ldots,q_{N_s}\right]^H$ with $q_{n}=e^{j\theta_{n}}$\;
$\mathbf{v}\gets \left[\mathbf{q}^T,t\right]^T$ with $|t|=1$\;
$\mathbf{V}\gets\mathbf{v}\mathbf{v}^H$\;}
\ForEach{Transmission Block}{
Estimate $\mathbf{f}$, $\mathbf{d}$, and $\mathbf{G}$\;
$\boldsymbol{\chi}\gets \mathrm{diag}(\mathbf{f})\mathbf{G}$\;
$\mathbf{C}\gets \begin{bmatrix}\boldsymbol{\chi}\boldsymbol{\chi}^H&\boldsymbol{\chi}\mathbf{d}^H\\ \mathbf{d}\boldsymbol{\chi}^H& \|\mathbf{d}\|^2\end{bmatrix} $\;
Solve \eqref{RIS_EQN_optimizationTrace} using CVX\;
Decompose $\mathbf{V}^\star=\mathbf{U}\boldsymbol{\Sigma}\mathbf{U}^H$\;
$\bar{\mathbf{v}}\gets \mathbf{U}\boldsymbol{\Sigma}^{1/2}\mathbf{r}$, where $\mathbf{r}\in \mathcal{CN}(\mathbf{0},\mathbf{I}_{N_s+1})$\;
Adjust RIS with $ \boldsymbol{\Phi}_{JT}^\star =\mathrm{diag}\left \{e^{j \arg\left( \left[\frac{\bar{\mathbf{v}}}{\bar{v}_{_{N_s+1}}} \right]_{1:N_s}\right)}\right\}$\;
All users jointly transmit\;
}
\end{algorithm}

\SetKwComment{Comment}{/* }{ */}
\RestyleAlgo{ruled}
\begin{algorithm}
\caption{Opportunistic RIS Transmission} \label{alg:IRS002}
\ForEach{Transmission Block}{
Estimate $\mathbf{f}$, $\mathbf{d}$, and $\mathbf{G}$\;
\ForEach{User $k$}{
$\mathbf{\Phi}^{\star}_{k}\gets e^{j\left( \arg\left( d_k\right)  -\arg\left( \mathrm{diag}(\mathbf{f})\mathbf{g}_k\right)\right)}$\;
$R_k^\star \gets \log  \left[  1+\frac{\left\|\mathbf{f} \boldsymbol{\Phi} \mathbf{g}_k +d_{k}\right\|^2P_u}{\sigma_n^2}  \right]$\;
}
$k^\star \gets \arg \max_{k=1,\ldots,K} \bigl(R_k^\star \bigr)$\;
$k^\star$ transmits while $k\neq k^\star$ turn off\;
}
\end{algorithm}

\subsection{Opportunistic Transmission}
If multiple users fade independently, the probability that one of the users experiences strong channel quality is substantially higher than that of a single user. Therefore, the sum capacity can be improved by exploiting the effect of multi-user diversity. By assigning the shared transmission resource only to the best user, the total throughput of the system is maximized. The more users the system can schedule, the stronger channel the best user probably has. Based on this observation, we propose an opportunistic scheme for the RIS-aided MU-MIMO system.

Accordingly, the \textit{single-user bound} is rewritten as
\begin{equation} \label{EQN_IRS_SingleUserBound}
  R_k < \log  \left[  1+\frac{\left\|\mathbf{f} \boldsymbol{\Phi} \mathbf{g}_k +d_{k}\right\|^2P_u}{\sigma_n^2}  \right],\: \forall\: k.
\end{equation}
The optimal phase-shift matrix is given by
\begin{equation} \label{eqnIRScomplexityQ}
    \mathbf{\Phi}^{\star}_{k}=\mathrm{diag}\left \{e^{j\left( \arg\left( d_k\right)  -\arg\left( \mathrm{diag}(\mathbf{f})\mathbf{g}_k\right)\right)}\right\}.
\end{equation}
Substituting \eqref{eqnIRScomplexityQ} into \eqref{EQN_IRS_SingleUserBound} yields the maximal achievable rate of user $k$, denoted by $R_k^\star$. The philosophy of the opportunistic transmission is to determine the best user with the largest rate, mathematically,
\begin{equation}
       k^\star=\arg \max_{k\in\{1,\ldots,K\}} \Bigl(R_k^\star\Bigr),
    \end{equation}
and then assigning the shared transmission resource merely to $k^\star$. Other users turn off while the best user transmits its signal.
The sum capacity of OT is computed by
\begin{align}   \nonumber
  C_{OT} & =  \max_{k\in\{1,\ldots,K\}} \left (\log  \left[  1+\frac{\left\|\mathbf{f} \boldsymbol{\Phi}_k^\star \mathbf{g}_k +d_{k}\right\|^2P_u}{\sigma_n^2}  \right]\right)\\
  &=\max_{k\in\{1,\ldots,K\}} \left ( \log\left(1+\frac{ \Bigl|\sum_{n=1}^{N_s}|f_n||g_{n,k}|+|d_k| \Bigr|^2 P_u}{\sigma_n^2} \right) \right).
\end{align}

\section{Numerical results}

\begin{figure}[!h]
    \centering
    \includegraphics[width=0.33\textwidth]{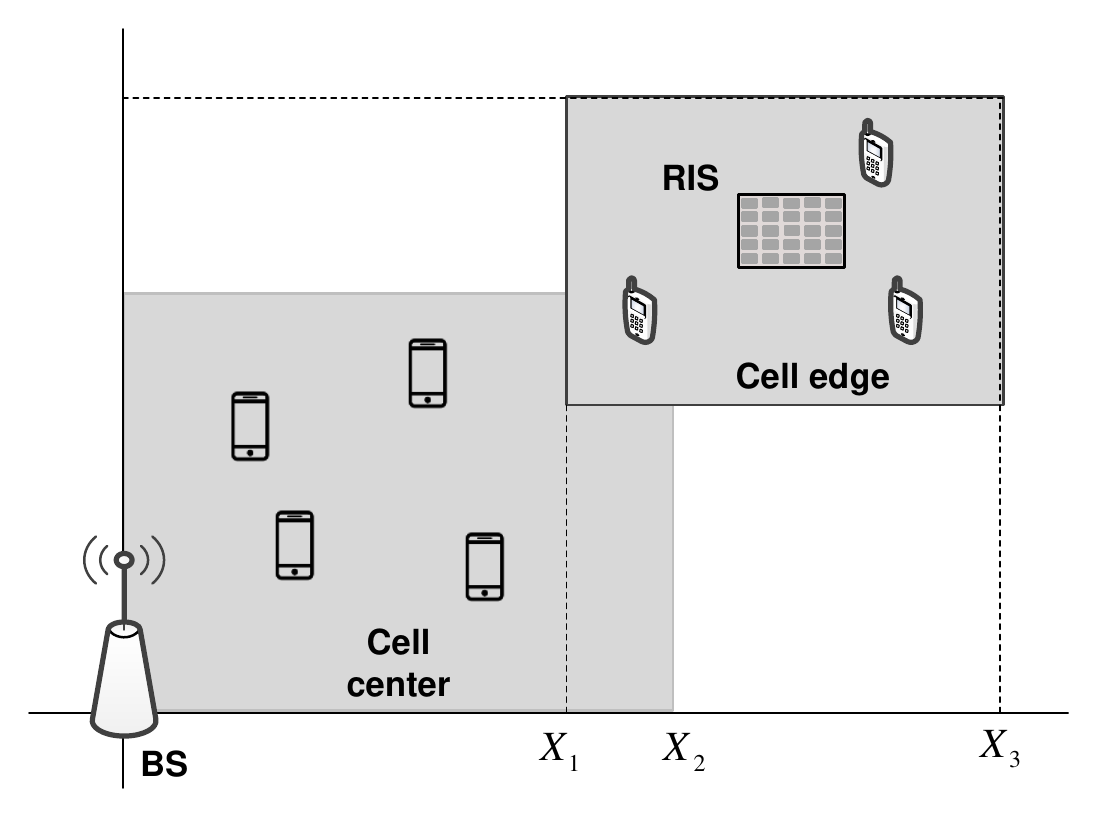}
    \caption{Simulation setup of a multi-user RIS system, where the cell coverage is comprised of a cell-center area and a cell-edge area. }
    \label{diagram:Simulation}
\end{figure}

Monte-Carlo simulations are conducted to evaluate the performance of joint and opportunistic transmission in an RIS-aided MU-MIMO system. This section first elaborates the simulation parameters and then provides some representative numerical results in terms of the sum throughput.
Without loss of generality, we established a simulation scenario as shown in \figurename \ref{diagram:Simulation}. The BS is located at the original point of the coordinate system, while the RIS with $N_s=200$ reconfigurable elements is deployed in the middle of the cell edge. Cell-center users distribute randomly over a square area with the side length of $X_2=\SI{300}{\meter}$, while cell-edge users distribute randomly over another square area from $X_1=\SI{250}{\meter}$ to $X_3=\SI{500}{\meter}$. The power constraint of the UE is assumed to be $P_u=1.0\mathrm{W}$ over a signal bandwidth of $1\mathrm{MHz}$.  The noise power density is $-174\mathrm{dBm/Hz}$ with the noise figure  $9\mathrm{dB}$.
The large-scale fading is distance-dependent, computed by
$\sigma_{c}^2=10^\frac{\mathcal{L}+\mathcal{S}}{10}$, where $\mathcal{L}$ denotes the path loss, and $\mathcal{S}\sim \mathcal{N}(0,\sigma_{sd}^2)$ is the Log-Normal shadowing with a standard derivation $\sigma_{sd}=8\mathrm{dB}$. The COST-Hata model (refer to \cite{Ref_jiang2021cellfree}) is employed to determine $\mathcal{L}$ using the break points of $10\mathrm{m}$ and $50\mathrm{m}$, the carrier frequency of $f_c=1.9\mathrm{GHz}$, the BS/RIS height of $15\mathrm{m}$, and the UE height of $1.65\mathrm{m}$. Due to the line of sight, the path loss of the BS-RIS channel can be calculated by $\mathcal{L}_0/d^{-\alpha}$,
where $\mathcal{L}_0=\SI{-30}{\decibel}$ is the path loss at the reference distance of \SI{1}{\meter}, the path-loss exponent $\alpha=2$, and the Rician factor $\Gamma=5$.

Our simulation provides a comprehensive comparison among different schemes, including: 1) the conventional direct communications (DC) where the UEs directly access to the BS without the aid of RIS; 2) FDMA that randomly selects a user for optimizing the RIS coefficients; 3) FDMA-US means exhaustive search to determine the best user for optimizing the RIS coefficients; 4) TDMA; 5) random phase shift (RPS) where the phase shifts of the RIS elements are \textit{randomly} set; 6) JT; 7) The upper bound of the JT, see \eqref{IRS_EQN_upper}; and 8) OT.
\begin{figure*}[!t]
\centerline{
\subfloat[]{
\includegraphics[width=0.43\textwidth]{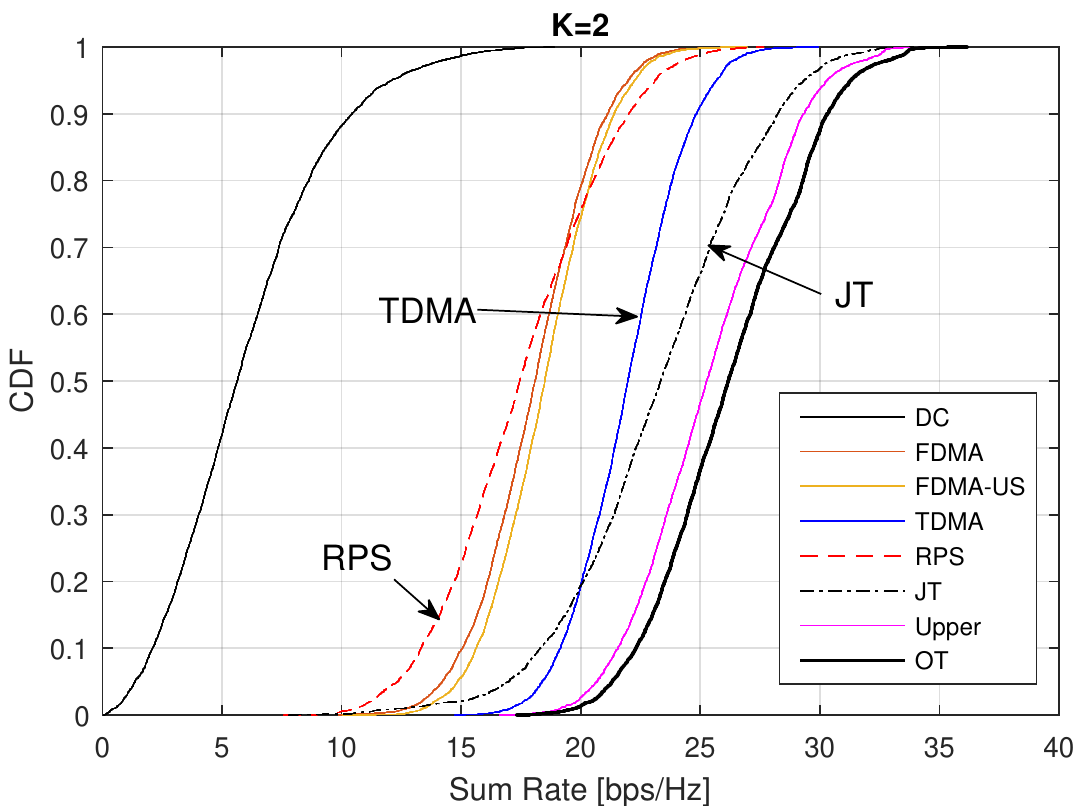}
\label{Fig_results1}
}
\hspace{5mm}
\subfloat[]{
\includegraphics[width=0.43\textwidth]{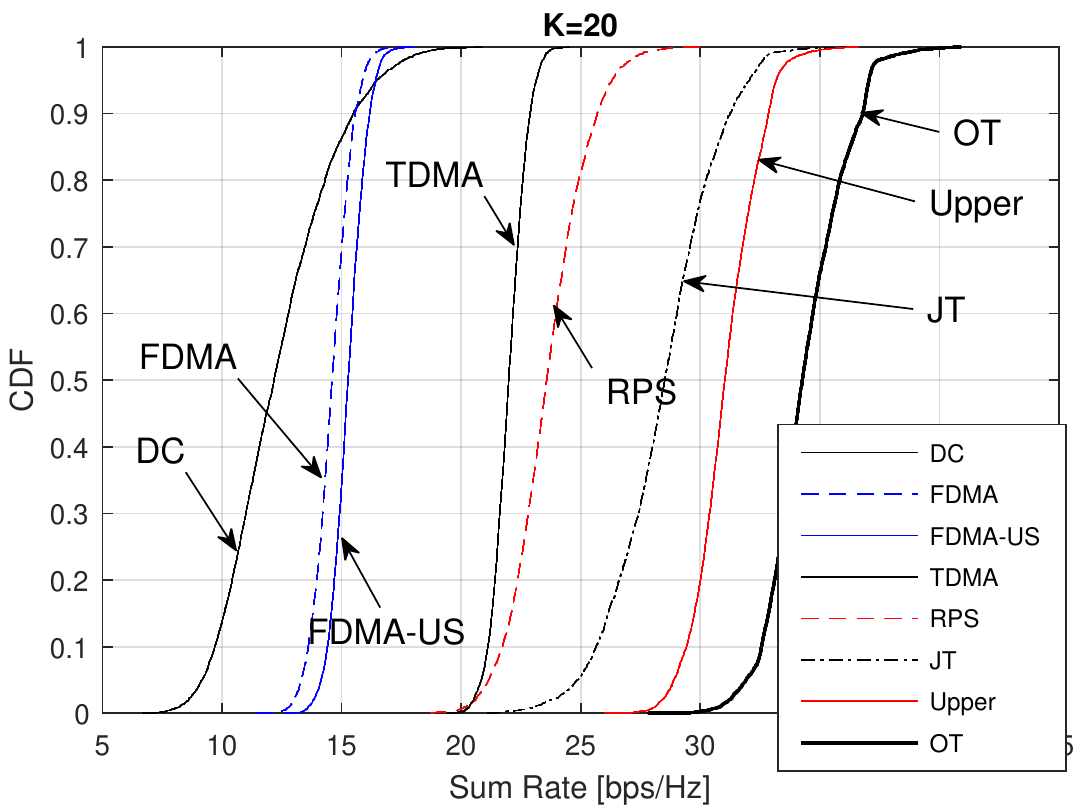}
\label{Fig_results2}
}}
\caption{Performance comparison of different transmission schemes in the uplink of a RIS-aided multi-user system: (a) CDFs in terms of the sum rate with two users, and (b) CDFs in terms of the sum rate with twelve users. }
\label{Fig_Result}
\end{figure*}

Cumulative distribution function (CDF) of the sum rate is employed as the performance metric. To provide some insights, we first  compare the CDFs of different schemes with the minimal number of $K=2$ users, consisting of a cell-center user and a cell-edge user. As shown in \figurename \ref{Fig_results1}, the conventional DC system achieves the $50\%$-likely or median sum rate of  \SI{5.6}{\bps\per\hertz^{}}. The deployment of RIS can substantially boost the system performance, where FDMA without and with user selection have a sum rate of around \SI{18.1}{\bps\per\hertz^{}} and \SI{18.5}{\bps\per\hertz^{}}, respectively. TDMA obviously outperforms FDMA with a $50\%$-likely sum rate of approximately \SI{22}{\bps\per\hertz^{}}. That is because the RIS can provide time-selective reflection dedicated for each TDMA user, whereas only the signal transmission of a single FDMA user can get aids owing to the lack of frequency-selective reflection.
As we expected, JT is superior to TDMA, where the $50\%$-likely rate is increased to approximately \SI{23.3}{\bps\per\hertz^{}}. That is because the signal transmission of each JT user fully exploits the time-frequency resource, in contrast to $1/K$ degree of freedom per OMA user. If the phase shifts are random, the result is \SI{17.4}{\bps\per\hertz^{}}, which justifies the effectiveness of the joint reflection optimization. The multi-user gain due to opportunistic user selection is solid, where OT has a median rate of \SI{26.2}{\bps\per\hertz^{}}, better than the upper performance bound of JT. In addition, we also illustrate the performance comparison in the case of $K=20$ users, as illustrated in \figurename \ref{Fig_results2}, where similar conclusions can be drawn from the numerical results. It is noted that the multi-user gain of OT becomes large with the increasing number of users.

\section{Conclusions}
Based on the insights provided by the capacity analysis, we proposed two novel schemes, i.e., \textit{joint transmission} and \textit{opportunistic transmission} for RIS-aided multi-user MIMO communications system. The superiority of the proposed schemes over the conventional orthogonal multiple access in terms of achievable sum rate was extensively justified through Monte-Carlo simulation. Particularly, opportunistic transmission has low complexity since it relies only on the best user selection, but obviously outperforming joint transmission, especially when the number of users becomes large.  Regardless of high complexity raised by the semidefinite relaxation of quadratically constrained quadratic program,  joint transmission still cannot compete with opportunistic transmission. The findings of this paper inspires us to further exploit multi-user diversity and opportunistic communications in RIS-aided systems.  on
\bibliographystyle{IEEEtran}
\bibliography{IEEEabrv,Ref_WCNC}

\end{document}